\documentstyle[12pt,cite]{article}
\def\opensquare{\thicklines\raisebox{-8pt}{\framebox(20,20){}}}
\def\sector#1#2{\ {\scriptstyle #1}\hspace{1mm}
\mathop{\opensquare}\limits_{\raisebox{-1mm}{$\scriptstyle#2$}}\hspace{1mm}}

\def\bfe#1{{{\bf e}\left[#1\right]}}

\begin{document}

\addtolength{\baselineskip}{.3mm} \thispagestyle{empty}
\vspace{-1.5cm}
\begin{flushright}
  KEK-TH-403\\
%{KEK preprint} ???\\
July 1994
\end{flushright}
\vspace{15mm}
\begin{center}
  {\large\sc{Elliptic Genera of $N=2$ Hermitian Symmetric Space
      Models}}\\[18mm]
  {\sc Toshiya Kawai}\footnote{E-mail address: kawai@theory.kek.jp}\\[8mm]
  {\it National  Laboratory for High Energy Physics (KEK),\\[2mm]
   Tsukuba, Ibaraki 305, Japan} \\[27mm]
\end{center}
\vspace{1.5cm}
\begin{center}
  {\sc Abstract}
\end{center}
\vspace{5mm}
\noindent

Expressions are given for the elliptic genera of the Kazama-Suzuki
models associated with hermitian symmetric spaces when the problems of
field identifications are absent. We use the models' known Coulomb gas
descriptions.

\vspace{2.5cm}
\begin{flushleft}
  {\tt hep-th/9407009}
\end{flushleft}

\newpage

Starting from the work of Witten \cite{rWitteni}\ the past year has
witnessed a resurgence of research activities in the subject of
elliptic genus \cite{rSW,rWittenii,rAKMW}\ in the particular context
of $N=2$ super(conformal) field theories
\cite{rDY,rKYY,rDAY,rHen,rAS,rBH,rKM,rNW}. Through this development
the $N=2$ elliptic genera have been computed for various models
including the minimal models, Landau-Ginzburg models and their
orbifolds, and sigma models.  It is natural to ask if we can extend
this list so as to cover another large class of $N=2$ superconformal
field theories, the so-called Kazama-Suzuki models \cite{rKS}. Of
course for some special cases of the models that admit Landau-Ginzburg
formulations we already have one of the expressions at hand, however,
for general Kazama-Suzuki models their elliptic genera have not been
obtained.

In this note we give one possible expression for the elliptic genera
of the Kazama-Suzuki models associated with hermitian symmetric spaces
(henceforth abbreviated as HSS models). The HSS models are known to
admit the Coulomb gas descriptions, {\it i.e.\/} those in terms of a
combined system of bosons and parafermions
\cite{rFLMW,rLerche,rEHY,rEKMY}\ and we will essentially make use of
this fact for our purpose. In particular we will adhere to the point
of view advocated in \cite{rEKMY}\ and some of the technologies
encountered there will be exploited here also.

Our starting point is a slight extension of the branching relation
considered in \cite{rKawai}\ as a Lie algebraic extension of Gepner's
branching relation for the $N=2$ minimal model \cite{rGepner0}.
Before writing this down we have to explain our notation.

We fix, once and for all, a simple Lie algebra $\bf g$ of rank $n$.
We use the convention in which the length of any long root is equal to
$\sqrt{2}$, thus $\theta^2=2$ where $\theta$ is the highest root of
$\bf g$.  The simple roots and fundamental weights of $\bf g$ are
denoted respectively by $\alpha_1,...,\alpha_n$ and
$\omega_1,...,\omega_n$.  The Weyl vector is defined to be
$\rho=\frac{1}{2}\sum_{\alpha\in\Delta^+}\alpha=\sum_{i=1}^n\omega_i$
where $\Delta^+$ is the set of positive roots of $\bf g$.  Given a
root $\alpha$ the corresponding coroot is written as
$\alpha^\vee=\frac{2}{\alpha^2}\alpha$.  We frequently use the
lattices $P=\sum_{i=1}^n {\bf Z}\omega_i$ (the weight lattice),
$Q=\sum_{i=1}^n {\bf Z}\alpha_i$ (the root lattice) and
$Q^\vee=\sum_{i=1}^n {\bf Z}\alpha_i^\vee ($the coroot lattice) as
well as the set $P_+^k =\{\lambda \in \sum_{i=1}^n {\bf Z}_{\geq
  0}\,\omega_i :\lambda \cdot \theta \leq k \}$. The dual Coxeter
number of $\bf g$ is denoted by $g$.  Let $W$ be the Weyl group of
$\bf g$. We write the length of $w\in W$ as $\ell(w)$.  The notation
$\bfe{x}=\exp (2 \pi ix)$ will be used throughout.

The theta function of level $k$ is defined by
\begin{equation}
  \Theta_{\mu,k}(\tau,u)=\sum\limits_{\beta\in Q^\vee+\mu/k}
  \bfe{\tau{\frac{k}{2}\beta^2}+ k u\cdot \beta}\,,\quad \mu\in
  P/kQ^\vee\,,
\end{equation}
and transforms as
\begin{equation}
  \Theta_{\mu,k}\left(-\frac{1}{\tau},\frac{u}{\tau}\right)=
  (-i\tau)^{\frac{n}{2}} \bfe{\frac{k}{2}\frac{u^2}{\tau}}
  \sum_{\mu'\in P/k Q^\vee}B_{\mu,\mu'}^{(k)}
  \Theta_{\mu',k}(\tau,u)\,,
\end{equation}
where
\begin{equation}
  B_{\mu,\mu'}^{(k)}
  =|P/kQ^\vee|^{-1/2}\bfe{-\frac{\mu\cdot\mu'}{k}}\,.
\end{equation}
The Weyl-Kac character at level $k$ of the affine Lie algebra $\hat
{\bf g}$ is
\begin{equation}
  \chi^\Lambda(\tau,u)=\frac{\sum_{w\in W}(-1)^{\ell(w)}
    \Theta_{w(\Lambda+\rho),k+g}(\tau,u)}{\sum_{w\in W}(-1)^{\ell(w)}
    \Theta_{w(\rho),g}(\tau,u)}\,,
\end{equation}
where $\Lambda\in P_+^k$ and transforms as
\begin{equation}
  \chi^\Lambda\left(-\frac{1}{\tau},\frac{u}{\tau}\right)=
  \bfe{\frac{k}{2} \frac{u^2}{\tau}} \sum_{\Lambda'\in
    P_+^k}A_{\Lambda,\Lambda'}^{(k)}\chi^{\Lambda'}(\tau,u)\,,
\end{equation}
with
\begin{equation}
  A_{\Lambda,\Lambda'}^{(k)}=i^{|\Delta^+|} \sum_{w\in
    W}(-1)^{\ell(w)} B_{w(\Lambda+\rho),\Lambda'+\rho}^{(k+g)}\,.
\end{equation}

The fundamental branching relation we consider is
\begin{equation}
  \chi^\Lambda(\tau,u)\Theta_{\mu,g}(\tau,u-v) =\sum\limits_{\nu\in
    P/(k+g)Q^\vee} \chi^{\Lambda,\mu}_\nu(\tau,v)
  \Theta_{\nu,k+g}\left(\tau,u - \frac{g}{k+g}v \right)\,,
\label{branch}
\end{equation}
which reduces to the one for the $N=2$ minimal models in the case
${\bf g}={\bf su(2)}$.  Using the multiplication formula
\begin{eqnarray}
  &&\Theta_{\lambda,m}(\tau,u)\Theta_{\mu,n}(\tau,v)=\nonumber\\[2mm]
  &&\hspace{5mm}\sum\limits_{\gamma\in
    Q^\vee/(m+n)Q^\vee}\Theta_{n\lambda - m\mu + mn\gamma, mn(m+n)
    }\left(\tau,\frac{u-v} {m+n}\right)
  \Theta_{\lambda+\mu+m\gamma,m+n}\left(\tau,\frac{mu+nv}{m+n}\right)\,,
\end{eqnarray}
it is straightforward to find
\begin{equation}
  \chi^{\Lambda,\mu}_\nu(\tau,v)=\sum\limits_{\gamma\in Q^\vee}
  c^\Lambda_{\nu-\mu+g\gamma}(\tau)\,
  \bfe{\frac{\tau}{2f}\left(\frac{g}{k+g}\nu-\mu+g\gamma\right)^2+
    {v}\cdot \left(\frac{g}{k+g}\nu-\mu+g\gamma\right)}\,,
\label{chiform}
\end{equation}
where $c^\Lambda_\lambda(\tau)$ is the string function defined through
$\chi^\Lambda(\tau,u)=\sum_{\lambda\in P/kQ^\vee}
c^\Lambda_\lambda(\tau) \Theta_{\lambda,k}(\tau,u)$ and
\begin{equation}
  f=\frac{kg}{k+g}\,.
\end{equation}

As an immediate consequence of (\ref{branch}), we find that
\begin{eqnarray}
  && \chi^{\Lambda,\mu+gQ^\vee}_{\nu+(k+g)Q^\vee}(\tau,v)=
  \chi^{\Lambda,\mu}_{\nu}(\tau,v) \label{chishiftprop}\\
  &&\chi^{\Lambda,\mu}_{\nu}(\tau+1,v)=
  \bfe{h^{\Lambda,\mu}_{\nu}-\frac{c_{\rm WZW}}{24}}
  \chi^{\Lambda,\mu}_{\nu}(\tau,v)\\
  &&\chi^{\Lambda,\mu}_{\nu}\left(-\frac{1}{\tau},\frac{v}{\tau}\right)=
  \bfe{\frac{f}{2}\frac{v^2}{\tau}}\sum_{\Lambda'\in
    P_+^k}\sum_{\mu'\in P/gQ^\vee} \sum_{\nu'\in
    P/(k+g)Q^\vee}S_{(\Lambda,\mu,\nu),(\Lambda',\mu',\nu')}\,
  \chi^{\Lambda',\mu'}_{\nu'}(\tau,v)\,,
\end{eqnarray}
where
\begin{equation}
  h^{\Lambda,\mu}_{\nu} = \frac{\Lambda\cdot (\Lambda+ 2\rho)-
    \nu^2}{2(k+g)}+\frac{\mu^2}{2g}\,, \qquad c_{\rm
    WZW}=\frac{k\mathop{\rm dim}{\bf g}}{k+g}\,,
\end{equation}
and
\begin{equation}
  S_{(\Lambda,\mu,\nu),(\Lambda',\mu',\nu')}^{(k)}=
  A^{(k)}_{\Lambda,\Lambda'} B^{(g)}_{\mu,\mu'}
  \left(B^{(k+g)}_{\nu,\nu'}\right)^*\,.
\end{equation}
It also easily follows from (\ref{chiform}) that
\begin{eqnarray}
  &&\chi^{\Lambda,\mu}_{\nu}(\tau,v+\kappa\tau+\lambda)=\nonumber \\
  &&\ \ \ \ \ \ \ \bfe{\lambda\cdot\left(\frac{g}{k+g}\nu-\mu\right)}
  \bfe{-\frac{f}{2}(\kappa^2\tau+2\kappa\cdot v)}
  \chi^{\Lambda,\mu-g\kappa}_{\nu-g\kappa}(\tau,v)\,,
  \label{chiprop}
\end{eqnarray}
for $\kappa,\,\lambda\in P$.

We also need to introduce the alternating sum
\begin{equation}
  {\cal I}^\Lambda_\nu(\tau,v)=\sum_{w\in W}(-1)^{\ell(w)}
  \chi^{\Lambda,w(\rho)}_{\nu}(\tau,v)\,,
\end{equation}
which transforms as
\begin{eqnarray}
  &&{\cal I}^\Lambda_\nu(\tau,v+1)=\bfe{h^{\Lambda,\rho}_{\nu}-
    \frac{c_{\rm WZW}}{24}}{\cal I}^\Lambda_\nu(\tau,v)
  \label{ItransT} \\ &&{\cal
    I}^\Lambda_\nu\left(-\frac{1}{\tau},\frac{v}{\tau}\right)
  =\bfe{\frac{f}{2}\frac{v^2}{\tau}}\sum_{\Lambda'\in P_+^k}
  \sum_{\nu'\in P/(k+g)Q^\vee}A^{(k)}_{\Lambda,\Lambda'} (-i)^{\vert
    \Delta^+\vert}\left(B^{(k+g)}_{\nu,\nu'}\right)^* {\cal
    I}^{\Lambda'}_{\nu'}(\tau,v)\,, \label{ItransS}
\end{eqnarray}
where we used
\begin{equation}
  \sum_{w\in W}(-1)^{\ell(w)}B_{w(\rho),\rho}^{(g)}=(-i)^{\vert
    \Delta^+\vert}\,.
\end{equation}

We will construct the elliptic genera of HSS models taking
$\chi^{\Lambda,\mu}_\nu(\tau,v)$ (or actually ${\cal
  I}^\Lambda_\nu(\tau,v)$) as basic building blocks.  Before
presenting the result, we need to make a digression on some
mathematical technicalities relevant in our argument. Since almost all
of these are gathered in appendices of \cite{rEKMY}, we shall be very
brief.  For a fuller explanation and notation the reader is advised to
refer to that paper.

Let $J=\{i\in \{1,2,\cdots,n\} : a_i=1\}$ where $a_i$'s are positive
integers such that $\theta=\sum_{i=1}^na_i\alpha_i$ and are tabulated
for instance in \cite{rEKMY}. The set $J$ is non-empty iff ${\bf
  g}=A_n$, $B_n$, $C_n$, $D_n$, $E_6$ or $E_7$ and we shall be
exclusively concerned with these cases. Given a $\#\in J$ then ${\bf
  g}_\#$ is defined to be the semi-simple Lie algebra obtained by
deleting the node $\#$ from the Dynkin diagram of ${\bf g}$ and we
remark that $\alpha_\#$ is always a long root. For each $\#\in J$ we
obtain a hermitian symmetric space ${\bf g}/{\bf m_\#}$ where ${\bf
  m_\#}$ is the reductive subalgebra such that ${\bf m_\#}\simeq {\bf
  g}_\#\oplus {\bf u(1)}$ and we shall set
$D_\#=\frac{1}{2}\mathop{\rm dim}_{{\bf R}} ({\bf g}/{\bf m_\#})$. Let
$W_\#$ be the subgroup of $W$ corresponding to the Weyl group of ${\bf
  g}_\#$ and $W^\#=\{w\in W : w^{-1}(\Delta_\#^+)\subset \Delta^+\}$
where $\Delta_\#^+$ is the subset of $\Delta^+$ correspnding to the
set of positive roots of ${\bf g}_\#$. Kostant's lemma states that
there is a unique decomposition of any $w\in W$ as $w=w'w''$ where
$w'\in W_\#$ and $w''\in W^\#$. Hence $\vert W^\#\vert=\vert
W\vert/\vert W_\#\vert$. For the elements of $W^\#$ we have
\begin{equation}
  \ell(\sigma)=(\rho-\sigma(\rho))\cdot\omega_\#\,,\quad \forall
  \sigma\in W^\#\,.
\label{length}
\end{equation}
Denoting the translation by $k\omega_\#$ as $t_{k\omega_\#}$ and the
longest element in $W^\#$ as $\tilde w^\#$, we set
\begin{equation}
  \gamma_{\#,k}=t_{k\omega_\#}\tilde w^\#\,.
\end{equation}
Then the $\gamma_{\#,k}$'s, as $\#$ runs through $J$, generate the
group of diagram automorphism $\Omega_k$ which is isomorphic to $P/Q$
regardless of the value of the nonnegative integer $k$. It follows
that $\gamma_{\#,g}(\rho)=\rho$ or equivalently
 \begin{equation}
   \rho-\tilde w^\#(\rho)=g\omega_\#\,.
 \end{equation}
 The following formula is worth noting:
\begin{equation}
  w(\Lambda+\rho)+ j(k+g)\omega_\#\equiv w(\tilde
  w^\#)^{-j}(\gamma_{\#,k+g})^j(\Lambda+\rho) \pmod{(k+g) Q^\vee}\,,
  \quad j\in {\bf Z}\,,
\end{equation}
where $\Lambda\in P_+^k$ and $w\in W$ and especially
\begin{equation}
  w(\rho)+ jg\omega_\#\equiv w(\tilde w^\#)^{-j}(\rho) \pmod{g
    Q^\vee}\,, \quad j\in {\bf Z}\,.
\label{DAequiv}
\end{equation}
Furthermore we have
\begin{equation}
  D_\#=\ell(\tilde w^\#)=g\omega_\#^2=2\rho\cdot\omega_\#\,.
\label{HSSdim}
\end{equation}

The existence of diagram automorphism is closely related to field
identifications \cite{rGepneri}. In fact from what we have explained
we can easily deduce that
 \begin{equation}
   S^{(k)}_{\gamma_\#(\Lambda,\mu,\nu),(\Lambda',\mu',\nu')}=
   S^{(k)}_{(\Lambda,\mu,\nu),(\Lambda',\mu',\nu')}\,,
\end{equation}
where
\begin{equation}
  \gamma_\#(\Lambda,\mu,\nu)=(\gamma_{\#,k}(\Lambda),\mu+g\omega_\#,
  \nu+(k+g)\omega_\#)\,,
\end{equation}
and
\begin{equation}
  \chi^{\gamma_{\#,k}(\Lambda), \mu +
    g\omega_\#}_{\nu+(k+g)\omega_\#}(\tau,v)=
  \chi^{\Lambda,\mu}_{\nu}(\tau,v)\,.
\end{equation}
In general if the group of diagram automorphism acts non-freely it is
difficult to construct modular invariant partition functions.  In the
following we simply assume that the group of diagram automorphism acts
without fixed points thus evading subtle problems.

Hence we may define
\begin{eqnarray}
  {\cal Z}(\tau,v)&=&\frac{1}{\vert P/Q \vert}
  \sum_{\Lambda,\tilde\Lambda\in P_+^k} \sum_{\nu\in
    P/(k+g)Q^\vee}N^{(k)}_{\Lambda,\tilde \Lambda} {\cal
    I}^\Lambda_\nu(\tau,v) \left( {\cal I}^{\tilde \Lambda}_\nu
  ( \tau,0) \right)^* \label{ZGdef} \\ &=&\frac{1}{\vert P/Q\vert}
  \sum_{\Lambda,\tilde\Lambda \in P_+^k} \sum_{w\in W}(-1)^{\ell(w)}
  N^{(k)}_{\Lambda,\tilde \Lambda}{\cal I}^\Lambda_{w(\tilde
    \Lambda+\rho)}(\tau,v)\,, \label{ZGexp}
\end{eqnarray}
where $N^{(k)}_{\Lambda,\tilde \Lambda}$'s are non-negative integers
such that they define a modular invariant partition function of the
WZW model at level $k$, {\it i.e.\/} it satisfies
\begin{eqnarray}
  && N^{(k)}_{\Lambda,\tilde \Lambda}=0 \quad \mbox{if
    $\frac{\Lambda\cdot(\Lambda+2\rho)-
      \tilde\Lambda\cdot(\tilde\Lambda+2\rho)}{2(k+g)}\equiv 0
    \pmod{\bf Z}$} \label{TinvofN}\\ &&N^{(k)}_{\Lambda,\tilde
    \Lambda}=\sum_{\Lambda',\tilde\Lambda'\in P_+^k}
  A^{(k)}_{\Lambda,\Lambda'} \left( A^{(k)}_{\tilde\Lambda,
    \tilde\Lambda'} \right)^* N^{(k)}_{\Lambda',\tilde \Lambda'}
  \label{SinvofN}\\ &&N^{(k)}_{\gamma(\Lambda),\gamma(\tilde
    \Lambda)}= N^{(k)}_{\Lambda,\tilde \Lambda}\,,\qquad
  \gamma\in\Omega_k\,.
\end{eqnarray}
In the step going from (\ref{ZGdef}) to (\ref{ZGexp}) we used
\cite{rKawai}
\begin{equation}
  {\cal I}^\Lambda_\nu(\tau,0)= \sum_{w\in
    W}(-1)^{\ell(w)}\delta^{[P/(k+g)Q^\vee]}_{\nu,w(\Lambda+\rho)}\,.
\label{indexrel}
\end{equation}
Using (\ref{ItransT}), (\ref{ItransS}), (\ref{TinvofN}) and
(\ref{SinvofN}) we can show that
\begin{equation}
  {\cal Z} \left( \frac{a\tau +b}{ c\tau +d}, \frac{v}{c\tau +d}
\right)= \bfe{ \frac{f}{2} \frac{ c v^2}{ c\tau +d}} {\cal Z}( \tau,
v)\,,\quad \pmatrix{ a&b\cr c&d\cr} \in SL(2,{\bf Z})\,.
  \label{modularG}
\end{equation}

Let us fix $\times\in J$ and consider the HSS model associated with
${\bf g}/{\bf m}_\times$.  The one third of its Virasoro central
charge is
\begin{equation}
  \hat c=\frac{kD}{k+g}\,,
\end{equation}
where we have set $D=D_\times$.  Now we come to the main assertion of
this letter: {\it the elliptic genus of the {\rm HSS} model is given
  by}
\begin{equation}
  Z(\tau,z)=\frac{1}{\vert W_\times\vert}{\cal Z}(\tau,\omega_\times
  z)\,,
\end{equation}
{\it so long as the diagram automorphism acts freely.} To confirm this
we check \cite{rKYY}\ the modular transformation property
\begin{equation}
  Z \left( \frac{a\tau +b}{ c\tau +d}, \frac{z}{c\tau +d} \right)=
  \bfe{ \frac{\hat c}{2}\frac{ c z^2}{ c\tau +d}} Z(\tau, z)\,,\quad
  \pmatrix{ a&b\cr c&d\cr} \in SL(2,{\bf Z})\,,
\label{modular}
\end{equation}
and the double quasi-periodicity
\begin{equation}
  Z(\tau,z+\lambda\tau+\mu)=(-1)^{\hat c(\lambda+\mu)}\bfe{-\frac{\hat
      c}{2} (\lambda^2\tau+2\lambda z)}Z(\tau,z)\,,\quad
  \lambda,\mu\in h{\bf Z}\,,
\label{periodicity}
\end{equation}
where $h$ is the least positive integer such that
\begin{equation}
  hg\omega_\times\in (k+g)Q^\vee\,.
\label{defpropofh}
\end{equation}
The modular transformation property (\ref{modular}) follows
immediately from (\ref{modularG}) since $D=g\omega_\times^2$.  As for
the double quasi-periodicity (\ref{periodicity}) it satisfies to check
\begin{equation}
  {\cal I}^\Lambda_{w(\tilde \Lambda+\rho)}(\tau,
  \omega_\times(z+\lambda\tau+\mu))=(-1)^{\hat c(\lambda+\mu)}
  \bfe{-\frac{\hat c}{2}(\lambda^2\tau+2\lambda z)} {\cal
    I}^\Lambda_{w(\tilde \Lambda+\rho)}(\tau,\omega_\times z)\,,
\end{equation}
for $\lambda, \mu\in h{\bf Z}$.  To prove this we first note that
\begin{equation}
  (-1)^{\hat c h}=(-1)^{Dh}\,,
\label{signrel}
\end{equation}
which follows from (\ref{defpropofh}) since
\begin{equation}
  {\bf Z}\ni\frac{hg}{k+g}\omega_\times\cdot\rho=\frac{hgD}{2(k+g)}\,.
\end{equation}
We then apply (\ref{chiprop}) and use, besides (\ref{defpropofh}) and
(\ref{signrel}), the properties presented earlier including
(\ref{chishiftprop}), (\ref{DAequiv}) and (\ref{HSSdim}) while taking
into account the fact that $\alpha_\times$ is a long root.

In order to complete our identification we have to make sure that
$Z(\tau,z)$ has the proper $\chi_y$ genus (a.k.a the Poincar{\' e}
polynomial).  The $\chi_y$ genus is related to the elliptic genus by
$Z(0,z)=y^{-\frac {\hat c}{2}}\chi_y$ \cite{rKYY}. In the present case
we have that
\begin{eqnarray}
  \chi_y&=&\frac{1}{\vert P/Q\vert\,\vert W_\times\vert}\sum_{w\in W}
  \sum_{\Lambda\in P_+^k} N_{\Lambda\Lambda}^{(k)}\,
  y^{\omega_\times\cdot
    w\left(\frac{g}{k+g}(\Lambda+\rho)-\rho)\right)+ \frac {\hat
      c}{2}}\\ &=&\frac{1}{\vert P/Q\vert}\sum_{\sigma\in W^\times}
  \sum_{\Lambda\in P_+^k} N_{\Lambda\Lambda}^{(k)}
  \,y^{Q_k(\Lambda,\sigma)}\,,
\end{eqnarray}
where
\begin{equation}
  Q_k(\Lambda,\sigma)=\ell(\sigma)+\frac{g}{k+g}\omega_\times \cdot
  ( \sigma (\Lambda + \rho)-\rho)\,,
\end{equation}
which is in accordance with the results in \cite{rLVW,rHT,rGepnerii}
\cite{rEKMY}. Here we have used Kostant's lemma, (\ref{length}) and
the property that $\omega_\times$ is stabilized by $W_\times$. Note
that with our definition of $h$ we have $hQ_k(\Lambda,\sigma)\in {\bf
  Z}$ as must be so.  The Witten index can be obtained either from
$\chi_{y=1}$ or $Z(\tau,0)$ using (\ref{indexrel}) as
\begin{equation}
  \frac{\vert W^\times\vert}{\vert P/Q\vert} \sum_{\Lambda\in
    P^k_+}N^{(k)}_{\Lambda\Lambda}\,.
\end{equation}
\medskip

A few remarks are in order.

\smallskip

{\noindent ({\romannumeral 1})} it is obvious that if we take ${\bf
  g}={\bf su(2)}$ then our formula of the elliptic genus reduces to
the one given in \cite{rKYY}\ for the $N=2$ minimal model\footnote{The
  elliptic genus of the diagonal $N=2$ minimal model was first given
  by Witten \cite{rWitteni}. In ref.~\cite{rDY} the elliptic genera
  were erroneously written down for the remaining non-diagonal cases
  ({\it i.e.\/} $D$ and $E$).  The expressions there, contrary to the
  authors' claim, do not satisfy the correct modular transformation
  laws since they have included only the diagonal entries of the
  modular invariants.}.  (Note that ${\bf g}_\times=\emptyset$,
$W=W^\times=\{1,-1\}$ and $\vert P/Q\vert=2$.)  \bigskip

{\noindent ({\romannumeral 2})} it is straightforward to compute the
elliptic genus of the theory orbifoldized by ${\bf Z}_h$. According to
the formula presented in \cite{rKYY}, it is given by
\begin{equation}
  Z_{\mbox{orb}}(\tau,z)=\frac{1}{h}\sum_{\alpha,\beta=0}^{h-1}
  (-1)^{D(\alpha+\beta+\alpha\beta)}\sector\beta\alpha(\tau,z)\,,
\end{equation}
where
\begin{equation}
  \sector\beta\alpha(\tau,z)=\bfe{\frac{\hat c}{2} \alpha\beta}
  \bfe{\frac{\hat c}{2}(\alpha^2\tau+2\alpha
    z)}Z(\tau,z+\alpha\tau+\beta)\,,
\end{equation}
and we used (\ref{signrel}).  This can also be rewritten as
\begin{eqnarray}
  &&Z_{\mbox{orb}}(\tau,z)=\frac{1}{\vert P/Q\vert\,\vert W_\times
    \vert} \frac{1}{h}\sum_{\alpha,\beta=0}^{h-1}
  \sum_{\Lambda,\tilde\Lambda\in P_+^k} \sum_{\nu\in P/(k+g)Q^\vee}
  \nonumber \\[2mm] && \hspace{1.5cm}\times
  \bfe{\frac{g}{2(k+g)}(D\alpha+2\omega_\times\cdot\nu)\beta}
  N^{(k)}_{\Lambda,\tilde\Lambda}\, {\cal
    I}^\Lambda_\nu(\tau,\omega_\times z) \left( {\cal I}^{\tilde
    \Lambda}_{\nu+\alpha g\omega_\times} ( \tau,0) \right)^*\,.
\end{eqnarray}

In summary, we have presented a candidate formula of the elliptic
genus of HSS model inspired by its Coulomb gas description and have
checked that it has the pertinent properties.

\vspace{1.5cm}

I am grateful to S.-K.~Yang for discussion.

\end{document}